\documentclass[a4paper,fleqn]{cas-dc}

\ExplSyntaxOn
\keys_set:nn { stm / mktitle } { nologo }
\ExplSyntaxOff

\usepackage[version=4]{mhchem} 
\usepackage{todonotes}
\usepackage{xspace}
\usepackage{booktabs}
\usepackage{multirow}
\usepackage{subfigure} 
\usepackage{siunitx}
\usepackage{courier} 

\usepackage{amssymb}

\newcommand{\ATMOS}{\texttt{ATMOS}\xspace}
\newcommand{\pyatmos}{\texttt{PyATMOS}\xspace}
\newcommand{\chemicalList}{\ce{O2}, \ce{H2O}, \ce{CH4}, \ce{CO2}, \ce{H2} and \ce{N2}}
\newcommand{\numatmospheres}{124,314\xspace} 

\usepackage{listings}
\definecolor{codegreen}{rgb}{0,0.6,0}
\definecolor{codegray}{rgb}{0.5,0.5,0.5}
\definecolor{codepurple}{rgb}{0.58,0,0.82}
\definecolor{backcolour}{rgb}{0.95,0.95,0.92}

\lstdefinestyle{mystyle}{
  backgroundcolor=\color{backcolour},   commentstyle=\color{codegreen},
  keywordstyle=\color{magenta},
  numberstyle=\tiny\color{codegray},
  stringstyle=\color{codepurple},
  basicstyle=\ttfamily\footnotesize,
  breakatwhitespace=false,         
  breaklines=true,                 
  captionpos=b,                    
  keepspaces=true,                 
  numbersep=5pt,                  
  showspaces=false,                
  showstringspaces=false,
  showtabs=false,                  
  tabsize=2
}
\graphicspath{{./}{figures/}}

\usepackage[authoryear]{natbib}

\def\tsc#1{\csdef{#1}{\textsc{\lowercase{#1}}\xspace}}
\tsc{WGM}
\tsc{QE}
\tsc{EP}
\tsc{PMS}
\tsc{BEC}
\tsc{DE}

\begin{document}
\let\WriteBookmarks\relax
\def\floatpagepagefraction{1}
\def\textpagefraction{.001}

\shorttitle{PyATMOS}

\shortauthors{A Chopra et~al.}

\title [mode = title]{PyATMOS: A Scalable Grid of Hypothetical Planetary Atmospheres}                      

%
\author[1,2]{A Chopra}[orcid=0000-0002-1488-5829]
\cormark[1]
\ead{aditya.chopra@anu.edu.au}

\author[3]{A.~Bell}[orcid=0000-0003-0365-4731] \cormark[1]
\author[4]{W.~Fawcett}[orcid=0000-0003-2596-8264] \cormark[1]
\author[5]{R.~Talebi} \cormark[1]
\author[6]{D.~Angerhausen}[orcid=0000-0001-6138-8633] \cormark[2]
\author[7]{A.G.~Baydin}[orcid=0000-0001-9854-8100] \cormark[2]
\author[8]{A.~Berea}[orcid=0000-0003-2813-8469] \cormark[2]
\author[9]{N.A.~Cabrol}[orcid=0000-0001-7520-4364] \cormark[2]
\author[10]{C.~Kempes}[orcid=0000-0002-1622-9761] \cormark[2]
\author[11]{M.~Mascaro}[orcid=0000-0002-4534-7484] \cormark[2]

\affiliation[1]{organization={University of Groningen, Groningen, Netherlands}}
\affiliation[2]{organization={The Australian National University, Canberra, Australia}}
\affiliation[3]{organization={Insight Edge Inc., Tokyo, Japan}}
\affiliation[4]{organization={Cavendish Laboratory, University of Cambridge, Cambridge, United Kingdom}}
\affiliation[5]{organization={Georgia Institute of Technology, Atlanta, USA}}
\affiliation[6]{organization={ETH Zürich, Zürich, Switzerland}}
\affiliation[7]{organization={University of Oxford, UK}}
\affiliation[8]{organization={George Mason University, Fairfax, USA}}
\affiliation[9]{organization={The SETI Institute Carl Sagan Center, California, USA}}
\affiliation[10]{organization={Santa Fe Institute, New Mexico, USA}}
\affiliation[11]{organization={Google Inc., Mountain View, California, USA}}

\cortext[cor1]{NASA Frontier Development Lab 2018 Participant}
\cortext[cor2]{NASA Frontier Development Lab 2018 Mentor}

\begin{abstract}
Cloud computing offers an opportunity to run compute-resource intensive climate models at scale by parallelising model runs such that datasets useful to the exoplanet community can be produced efficiently. To better understand the statistical distributions and properties of potentially habitable planetary atmospheres we implemented a parallelised climate modelling tool to scan a range of hypothetical atmospheres.Starting with a modern day Earth atmosphere, we iteratively and incrementally simulated a range of atmospheres to infer the landscape of the multi-parameter space, such as the abundances of biological mediated gases (\ce{O2}, \ce{CO2}, \ce{H2O}, \ce{CH4}, \ce{H2}, and \ce{N2}) that would yield `steady state' planetary atmospheres on Earth-like planets around solar-type stars. Our current datasets comprises of \numatmospheres simulated models of exoplanet atmospheres and is available publicly on the NASA Exoplanet Archive. Our scalable approach of analysing atmospheres could also help interpret future observations of planetary atmospheres by providing estimates of atmospheric gas fluxes and temperatures as a function of altitude. Such data could enable high-throughput first-order assessment of the potential habitability of exoplanetary surfaces and sepcan be a learning dataset for machine learning applications in the atmospheric and exoplanet science domain.
\end{abstract}




\maketitle

\section{Introduction}
Understanding the nature and distribution of habitable environments in the universe, and the life forms that may inhabit them, is increasingly part of the primary science goals of remote and \textit{in situ} planetary exploration missions. Strategies \citep{NASEM2018,NASEM2019} and roadmaps \citep{DesMarais2008,Achenbach2015,Horneck2016} all suggest that identifying, exploring, and characterising extraterrestrial environments for habitability and biosignatures will be a key focus of planetary science research endeavors in the coming decade. Remote spectroscopy allows us to infer the atmospheric composition of exoplanets, but it will be challenging for even the latest generation of ground- and space-based telescopes to characterise the surface habitability of an exoplanet \citep{Robinson2018}. Additionally, while visible and infrared spectroscopy can help quantify the abundances of atmospheric gases such as \ce{O2}, \ce{H2O}, \ce{CO2}, \ce{CH4} and \ce{CO}, other gases such as \ce{H2}  and \ce{N2} have no permanent dipole moment and are difficult to quantify through spectroscopic observations \citep{Kaltenegger2017,Woitke2021}. 

All these gases collectively have significant control on the oxidation state of the atmosphere and the extent of disequilibrium in an atmosphere which is available to any potential surface biochemistry. Until ground-based Extremely Large Telescopes and or space-based mission concepts like HabEx, LUVOIR, or LIFE come to fruition, the expected SNRs associated with observations in the JWST-era are unlikely to be able to place strong constraints on the atmospheric compositions of exoplanets in circumstellar habitable zones \citep{Seager2017,Fujii2018,Kaltenegger2020a}. Modelling of explanatory atmospheres with limited observational constraints will remain \textit{modi operandi} of planetary habitologists for the foreseeable future. 

In an effort to more holistically understand the nature of potentially habitable atmospheres, we designed a modelling framework that allows concurrent simulation of hundreds of thousands of planetary atmospheres so that it would become possible to undertake `parameter sweeps' in a high-dimensional parameter space. Here we present the \pyatmos dataset, and the associated scalable modelling framework produced as part of the 2018 NASA Frontier Development Lab to explore the parameter space of planetary atmospheres that are conducive to habitable conditions on the planetary surface.

The universe is filled with stars similar to our Sun \citep{Robles2008} and exoplanet statistics suggest that rocky planets similar to our Earth are common \citep{Burke2015,Petigura2013,Bovaird2015,Hsu2019,Bryson2020}. Water, heat, chemical disequilibria, and energy sources would have been present on early wet rocky planets because of the universal nature of the processes that produced them \citep{Chopra2018,Lineweaver2012AnnRev}.

Since all life on Earth needs liquid water during some part of its life cycle, and the surface of the Earth is covered with it, the presence of liquid water on a planet's surface is taken as a necessary (but not sufficient) condition for life \citep{Lineweaver2012,McKay2014}. Even if water is a constituent of the initial inventory of volatiles on rocky planets in the circumstellar habitable zones of their host stars, surface liquid water can exist only within the relatively narrow range of pressures and temperatures and thus may be only a transient feature of most habitable planets \citep{Lineweaver2018,Chopra2016}. Thus, the search for extra-terrestrial life on exoplanets is in a large part a search for extra-terrestrial surface pressures and temperatures that are conducive to liquid water. The pressure and temperature on a planetary surface is in large part a function of the properties of the atmosphere above the surface.

The exoplanetary science community has been studying factors that can influence surface habitability of exoplanets such as surface temperatures, densities, compositions, tectonic regimes, atmospheric chemistry, and albedos \citep{Kasting2003,Gaidos2005,Nisbet2007,Zahnle2007,Lammer2009,Kopparapu2013,Seager2013,Cockell2016,Godolt2016,Kaltenegger2017,Boutle2017,Meadows2018a,Kite2018,Keles2018}. When it comes to remote detection in the near future, our search for life on potentially habitable planets will almost exclusively depend on our ability to spectrally characterise and understand the abiotic and potentially biotic contributions to atmospheric chemical disequilibria \citep{Kasting2009,Krissansen-Totton2018a,Seager2010,Vazquez2010}. If we are to find an unambiguous biosignature that can be remotely detected, and design instruments to detect them, we need to identify the range of atmospheres that should be priority targets for future observations \citep{Lovelock1965,Seager2017,Meadows2018,Schwieterman2018,Catling2018,Kiang2018,Walker2018}. We will also need to understand about what type and extent of biology could support, or at least be compatible with, the different atmospheres that could exist on exoplanets.

Planets within our solar system have strikingly different surface conditions, in large part because of the composition of the atmospheres they host. The next generation of telescopes will have the sensitivity required to determine the composition of exoplanetary atmospheres \citep{Fujii2018,Wang2018,Venot2018}. Remotely assessing the potential for life on the surface of a planet will require us to estimate the surface pressure and temperature to assess the likelihood of surface liquid water. The parameter space of possible atmospheres on exoplanets is large and exploring it is computationally intensive. 

In order to investigate such a large parameter space, we created a `set \& forget' workflow to run planetary atmosphere models within a scalable framework. To test the framework, we simulated a wide distribution of atmospheric compositions by varying six input parameters of the model. The six parameters varied were the concentrations gases: \ce{O2}, \ce{CO2}, \ce{H2O}, \ce{CH4}, \ce{H2}, and \ce{N2}. The gases were chosen because they are the most abundant gases in Earth's atmosphere and thus likely to be the gases whose concentrations will be of interest to future observations of potentially habitable exoplanets. Additionally, the surface fluxes of these gases have been biologically mediated by life on Earth through different metabolisms ever since the emergence of life on Earth \citep{Nealson1999,Nisbet2001}. Thus, studying  atmospheres with different concentrations and fluxes of these gases can not only better enable us to evaluate the surface habitability of potentially inhabited exoplanets but also inform estimates of the likelihood and type of life being present on a remotely characterised exoplanet. Our approach will help transition from a zero-dimensional model of a circumstellar habitable zone to a more nuanced Gaussian distribution which can parametrise the extent of habitability \citep{Lineweaver2018}.

\section{Method}

\subsection{Simulation of atmospheres with ATMOS} \label{sec:atmos}
To scan the parameter space of atmospheres, we employed the \ATMOS software package~\citep{Arney2016,Meadows2016} on a massively parallelised cloud-based process to create a database of exoplanet atmospheres.

The \ATMOS package, a coupled photochemistry-climate model\footnote{ \url{https://github.com/VirtualPlanetaryLaboratory/atmos}}, considers a 1-D column of gas through the atmosphere. It is configurable with input parameters such as the concentration or surface fluxes of different species of gases, the stellar type of the planet's host star, the gravitational field strength of the planet, and the distance between the planet and the host star. The output of \ATMOS is a 1-D column of the resultant atmosphere's temperature, pressure, gas concentrations and gas fluxes as a function of altitude.

\ATMOS uses a photochemical model to calculate the effect of UV radiation on the different gas species, and a climate model to calculate the temperature and pressure profile, as a function of altitude, of the different gases.

The photochemical model includes particle microphysics and is run first to generate an initial atmospheric state based on user-specified boundary conditions (gas mixing ratios and fluxes, the temperature-pressure profile and the incident stellar spectrum). For our analyses, we started with planetary boundary conditions set to the present-day Earth and stellar parameters set to the present-day Sun. Output files from the photochemical model for altitude, pressure and gas mixing ratios are then passed into the climate model as its initial conditions and the climate model runs until it reaches a converged state. The climate model then feeds updated temperature and gas profiles back into the photochemical model. The models iterate back and forth in this manner until convergence is reached\footnote{For the development history and details on the coupling and convergence of the two models, see~\citet{Arney2016,Meadows2016}}. \ATMOS can thus be described as a coupled set of differential equations, and the software works to find a local `steady state' solution for a given set of gas concentrations and fluxes as a function of altitude. A consequence of this is a strong dependence on the initial `seeded' state of atmospheric concentrations. The software can only solve the set of differential equations provided that the next set of initial conditions is not too far from that of the previous set of initial conditions, and therefore one must take small steps in parameter space to get from one set of gas concentrations to another. The increments were determined empirically in past usage of this code by \citet{Arney2016}. 

\begin{table}[!htb]
    \centering
    \begin{tabular}{cclc}
\toprule
Gas & Scan range & Increment & Modern Earth\\ \midrule
\multirow{2}{*}{\ce{O2}} & 0.0--0.3 & 0.02 & \multirow{2}{*}{0.21} \\
                         & 0.3--1.0 & 0.05 & \\ 
\multirow{2}{*}{\ce{CO2}} & 0.0--0.1 & 0.01 & \multirow{2}{*}{4.00 $\times$ 10\textsuperscript{$-$4}}\\
                          & 0.1--1.0 & 0.05 & \\    
\multirow{1}{*}{\ce{H2O}} & 0.0--0.9 & 0.05 & 1.23 $\times$ 10\textsuperscript{$-$ 2} \\ 
\multirow{1}{*}{\ce{CH4}} & 0.0--0.1 & 0.005 & 1.63 $\times$ 10\textsuperscript{$-$ 6} \\ 
\multirow{1}{*}{\ce{H2}} & 0.0--10\textsuperscript{$-$7} &  10\textsuperscript{$-$9} & 8.13 $\times$ 10\textsuperscript{$-$8}  \\ 
\ce{N2}+trace gases & --- & --- & 0.78 \\
\bottomrule     
    \end{tabular}
    \caption{Fractional scan range and increments of gases varied in order to explore the parameter space of atmospheres. We note that \ce{N2} was not varied in a step-wise manner as was done with the other gasses but was instead used to `fill' the remainder of the atmosphere if the combination of other gas concentrations did not add to 100\%. Trace gases were not varied and included as a fixed portion of the atmosphere.} \label{table:scanranges}
\end{table}

Table~\ref{table:scanranges} contains the list of scan ranges and increments which correspond to the step sizes, and the initial conditions corresponding to Modern Earth. The gas concentrations chosen to vary were \chemicalList. Other trace gases (including \ce{O3}) important to the composition of Earth's current atmosphere were incorporated into the models at Modern Earth concentrations, and not varied between the scans. Starting with a present-day Earth atmosphere, we iteratively and incrementally sampled atmospheres with different gas mixtures.

In order to explore the parameter space of atmospheric concentrations in a systematic manner, \pyatmos was configured to iteratively use previous atmosphere solutions that were within the defined increments of  the `target' conditions as `seeds'. A finished run would then go on to seed the initial state for a subsequent run, which would solve the state for some small permutation in each gas relative to the previous state. The `origin' state for the whole search strategy was defined by a \texttt{Modern Earth} template (a complete set of parameters corresponding to the present-day atmosphere of the Earth) and subsequent runs computed the atmospheric profiles in a parameter space `similar' to \texttt{Modern Earth}. The process would repeat until either the model run timed-out or the defined parameter space was scanned.

\subsection{Software and compute environment}

The \ATMOS software exhibits platform dependencies, in part attributable to its legacy piece-wise development in \texttt{Fortran}. To streamline the \ATMOS runs and maintain cross-platform consistency, we created a Docker image of \ATMOS based on the Ubuntu Linux distribution. This image guaranteed consistent performance on all host platforms. To automate the process of configuring \ATMOS for individual runs, we wrote a package called \pyatmos{} in \texttt{Python 3} (chosen for its flexibility, extensive community-driven resources and potential for further development by end-users). \pyatmos{} allows one to easily configure \ATMOS, run it, and extract the relevant results. 

\begin{figure}[!hbt]
    \centering
    \includegraphics[width=\hsize]{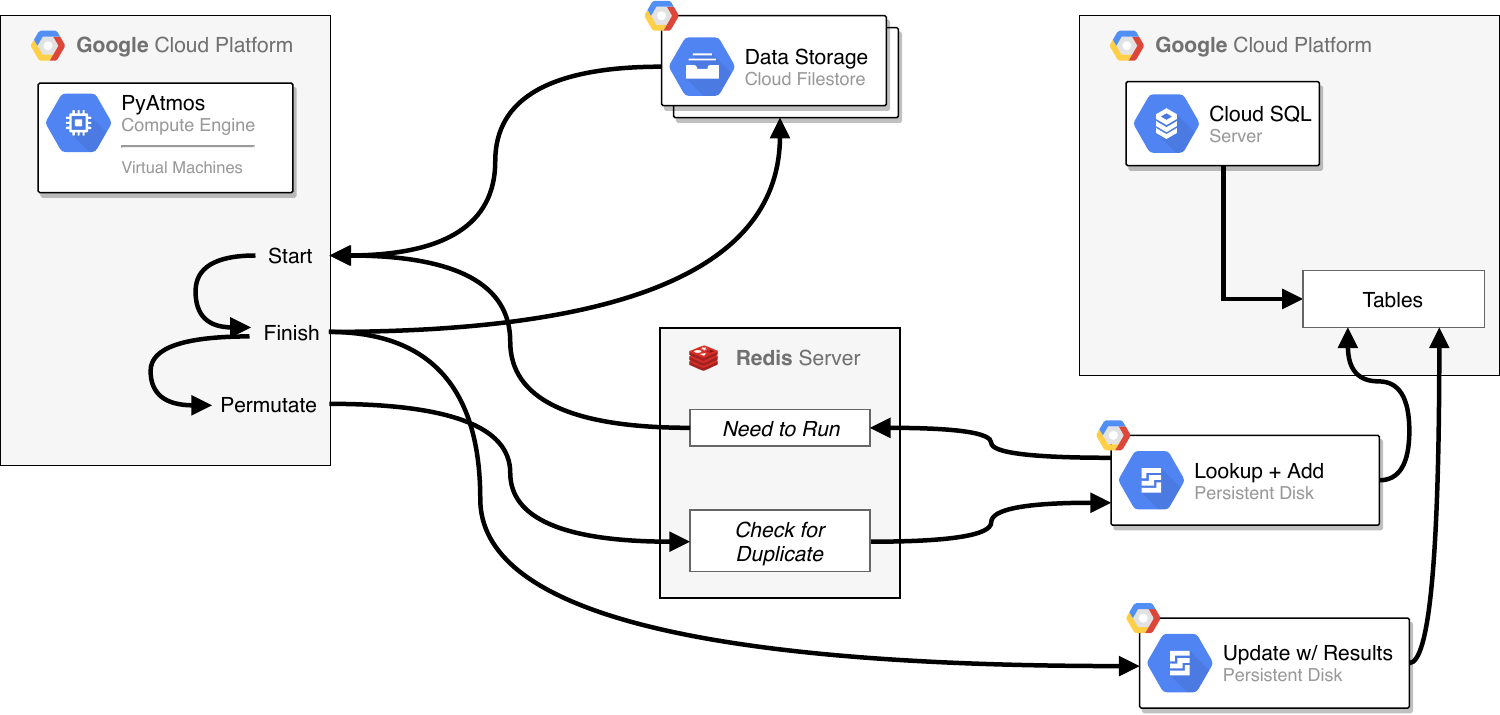}
    \caption{Cloud-computing work-queue for exploring a large parameter space of atmospheres.}
    \label{fig:technical}
\end{figure}

A Docker image loaded with \pyatmos, which inherited the original \ATMOS image, was created to instantiate thousands of individual cloud instances, all of which worked in parallel to search the atmospheric parameter space. Additional Python scripts were written to supervise a work-queue and designed to manage the run-constraints of \ATMOS outlined in Section~\ref{sec:atmos}. The work-queue is visualised in Fig.~\ref{fig:technical}.

The cloud instances spawned off thousands of identical virtual environments to compute the individual atmospheric concentrations with \pyatmos. Google Cloud Storage hosted all the data output by each run, and a SQL server stored a parsed log of all completed and queued runs. A Redis server tracked the completion of runs and allocated new work to each virtual machine.

Listing~\ref{code:pyatmos} shows how a series of gas concentrations is input to \pyatmos, the code is then run, and the results are stored in a specified output directory. Since \ATMOS requires a previously found stable atmosphere to `step' from in order to perform the new calculation, we  set the \textit{previous\_photochem\_solution} and \textit{previous\_clima\_solution} parameters of the \textit{atmos.run} function to strings containing the path to the relevant previous solution.

\begin{lstlisting}[language=Python, label={code:pyatmos}, caption={Simple example of running \pyatmos. A set of gas concentrations as inputs are run in a single iteration of \ATMOS with \pyatmos. When executed via a batch scheduling script, the same code enables parameter sweeps across a range of atmospheres.}]
import pyatmos
atmos = pyatmos.Simulation(
    docker_image = "registry.gitlab.com/frontierdevelopmentlab/astrobiology/pyatmos")
# setup the docker container    
atmos.start()
# Configuration for ATMOS
concentrations = {'H2O': 0.2, 'CO2': 0.0004, 'CH4': 1.63e-06, 'O2': 0.2, 'H2': 8.13e-08}
args = {
'species_concentrations' : concentrations,
        'output_directory' : "/home/results/"}
# Run the code
atmos.run(**args)
# Close the docker container
atmos.close()
\end{lstlisting}

\section{Results}

\begin{figure*}[!thb]
    \centering
    \includegraphics[width=\hsize]{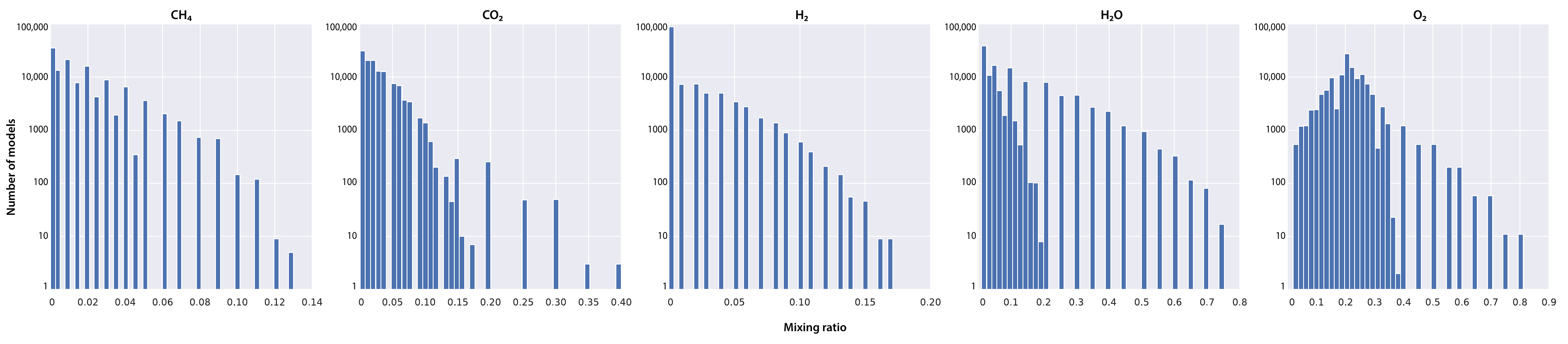}
    \caption{Histograms of input concentrations for \ce{CH4}, \ce{CO2}, \ce{H2}, \ce{H2O} and \ce{O2}. }
    \label{fig:input}
\end{figure*}

\begin{figure*}[!thb]
    \centering
    \includegraphics[width=\hsize]{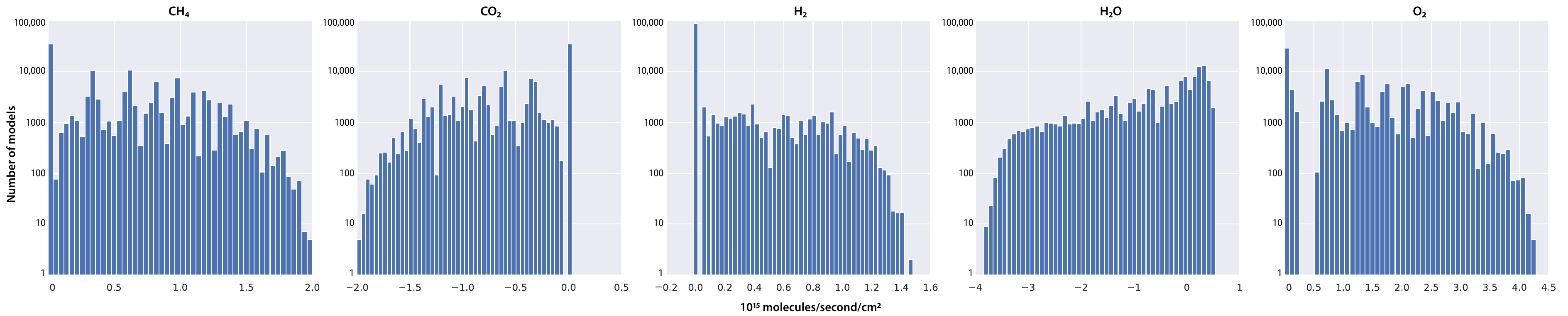}
    \caption{Histograms of output surface fluxes for \ce{CH4}, \ce{CO2}, \ce{H2}, \ce{H2O} and \ce{O2}. }
    \label{fig:output}
\end{figure*}

\begin{table*}[!htb]
    \centering
    \begin{tabular}{cccl}
\toprule
Column Name & Table Label & Units & Description \\ \midrule
input\_CH4          & Input CH4 concentration & fractional      & \ce{CH4} concentration at planet surface input to model\\
input\_CO2          & Input CO2 concentration & fractional      & \ce{CO2} concentration at planet surface input to model \\
input\_H2           & Input H2 concentration  & fractional      & \ce{H2} concentration at planet surface input to model \\
input\_H2O          & Input H2O concentration & fractional      & \ce{H2O} concentration at planet surface input to model \\
input\_O2           & Input O2 concentration  & fractional      & \ce{O2} concentration at planet surface input to model \\

concentration\_CH4  & CH4 Concentration       & fractional      & \ce{CH4} concentration at planet surface* \\
concentration\_CO2  & CO2 Concentration       & fractional      & \ce{CO2} concentration at planet surface* \\
concentration\_H2   & H2 Concentration        & fractional      & \ce{H2} concentration at planet surface* \\
concentration\_H2O  & H2O Concentration       & fractional      & \ce{H2O} concentration at planet surface* \\
concentration\_O2   & O2 Concentration        & fractional      & \ce{O2} concentration at planet surface* \\

flux\_CH4           & CH4 Flux                & \si{molecules\per\second\per\cm\squared} & \ce{CH4} flux required to maintain concentration at planet surface* \\
flux\_CO2           & CO2 Flux                & \si{molecules\per\second\per\cm\squared} & \ce{CO2} flux required to maintain concentration at planet surface* \\
flux\_H2            & H2 Flux                 & \si{molecules\per\second\per\cm\squared} & \ce{H2} flux required to maintain concentration at planet surface* \\
flux\_H2O           & H2O Flux                & \si{molecules\per\second\per\cm\squared} & \ce{H2O} flux required to maintain concentration at planet surface* \\
flux\_O2            & O2 Flux                 & \si{molecules\per\second\per\cm\squared} & \ce{O2} flux required to maintain concentration at planet surface* \\ 

pressure\_bar       & Pressure                & \si{\bar}       & Pressure at planet surface* \\
temperature\_kelvin & Temperature             & \si{\kelvin}    & Temperature at planet surface* \\
\bottomrule
\end{tabular}
    \caption{Column definitions of the summary table which contains \numatmospheres rows, where the data in each row summarises the varied input parameters, and the resulting output parameters (indicated by * in the description) for each atmosphere model.} \label{table:ColumnDefinitions}
\end{table*}

A total of \numatmospheres atmospheres were simulated and Fig.~\ref{fig:input} shows the distribution of concentrations sampled in this study. Although this represents only a small fraction of the 6-D parameter space possible within the scan constraints (limited by our compute-resource limits), the resulting dataset could be expanded for future studies. For each atmosphere, the temperature, pressure, gas concentration and gas fluxes were calculated as a function of altitude from 0--80\,km in 100 steps (a mean step-size of approximately 800\,m normally distributed with a standard deviation of 118\,m). The concentrations and fluxes for each of the gases listed in Table~\ref{table:scanranges} were calculated, along with several other trace gases with concentrations less than 1\,\%. Figure~\ref{fig:output} shows the distribution of surface fluxes for the five gases which were varied as input parameters in the atmosphere simulations.

The data from the simulated atmospheres are available to the community on the NASA Exoplanet Archive\footnote{https://exoplanetarchive.ipac.caltech.edu/cgi-bin/FDL/nph-fdl?atmos}. The interactive portal enables users to filter, preview, and download one or more models of interest. Table~\ref{table:ColumnDefinitions} describes the summary table which shows the input concentrations and the output parameters.

\begin{figure}[!hbt]
    \centering
    \includegraphics[width=0.8\hsize]{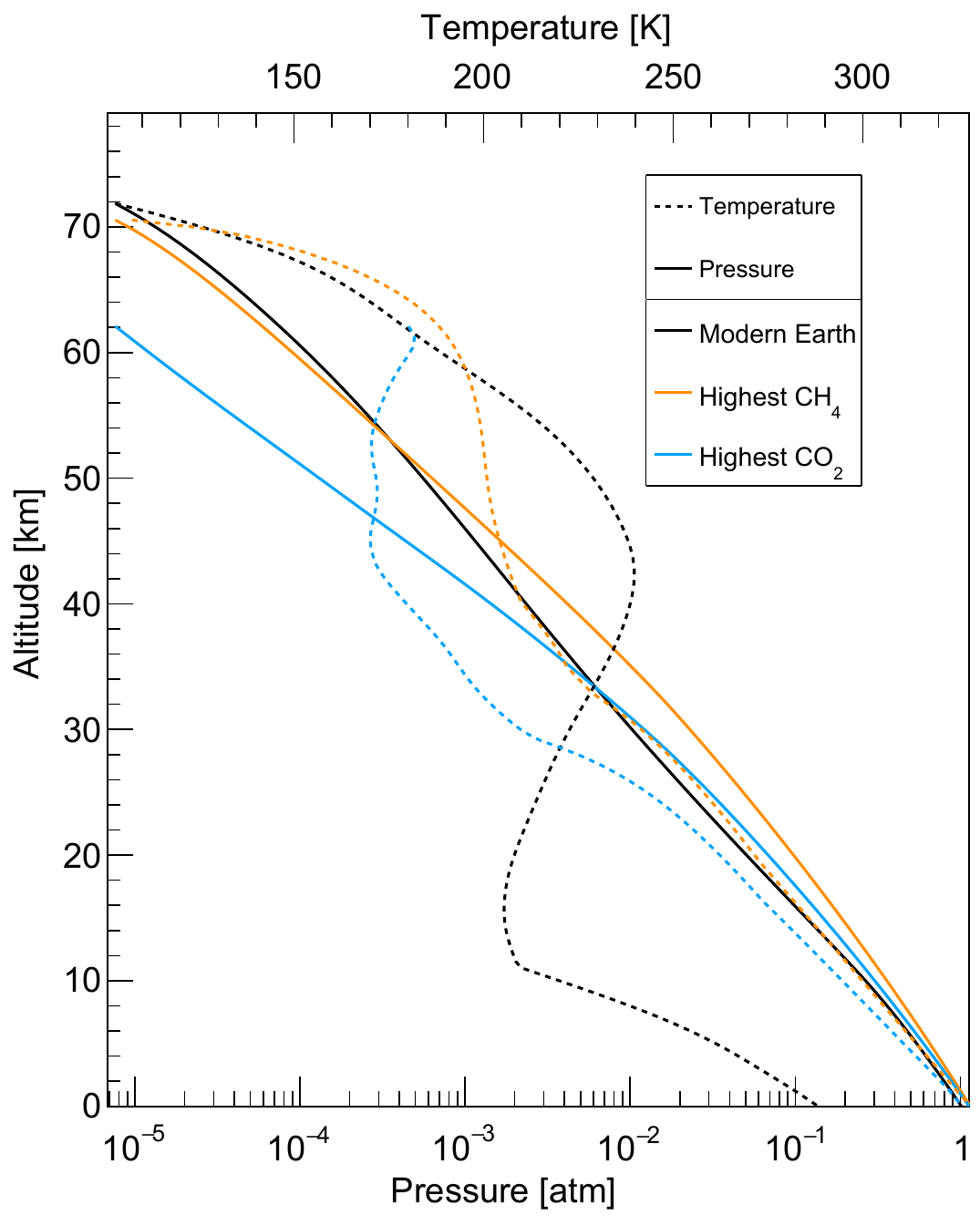}
    \caption{Temperature (dashed lines) and pressure (solid lines) profiles for three of the simulated atmospheres. 
    The black lines are for the `Modern Earth' atmosphere, the orange and blue lines correspond to the atmospheres simulated in this study with the largest concentrations of \ce{CH4} (13\%) and \ce{CO2} (40\%) respectively. }
    \label{fig:multi_planet}
\end{figure}

\begin{figure}[!hbt]
    \centering
    \includegraphics[width=0.8\hsize]{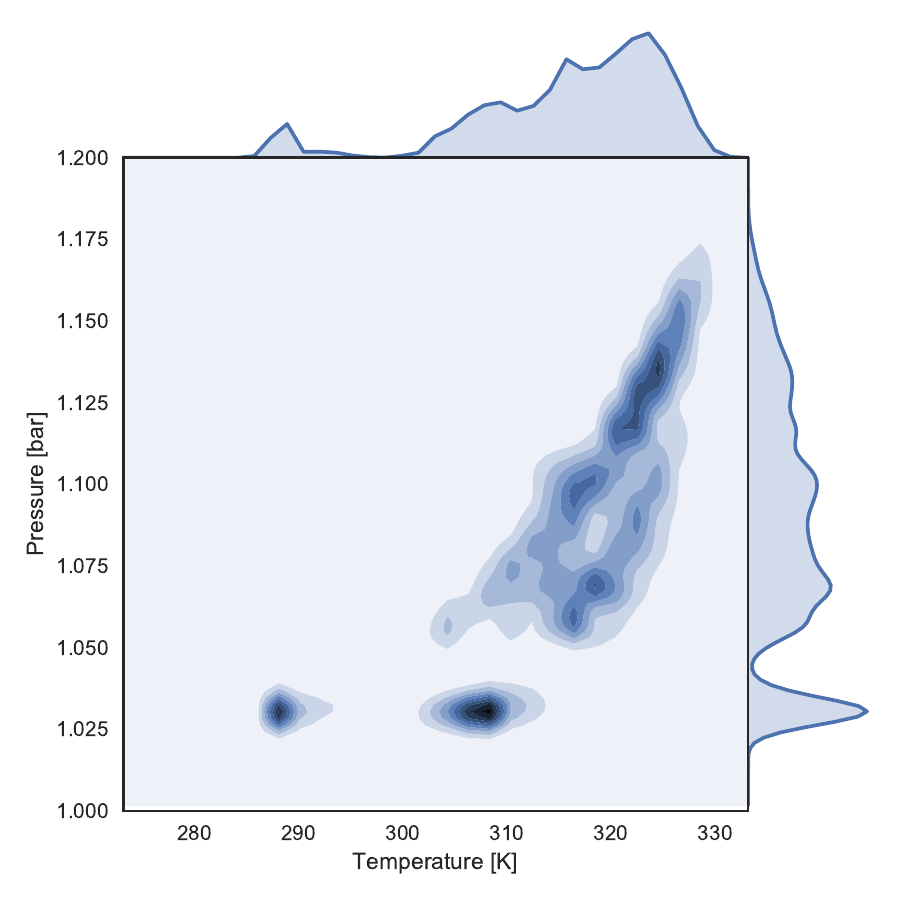}
    \caption{Distribution of atmospheres in the temperature versus pressure plane. The graphs above the upper and right axes are the 1-D density profiles of temperature and pressure, respectively. Although the distribution is sensitive to the priors applied to the scan and the total number of atmospheres scanned, similar analyses with larger datasets could help infer the frequency of different classes of habitable exoplanetary atmospheres and enable interpretation of biosignatures.}
    \label{fig:atmdist}
\end{figure}

Figure~\ref{fig:multi_planet} shows an example of three temperature- and pressure-profiles: the present-day Earth, the atmosphere with the largest concentration of \ce{CO2} (40\%), and the atmosphere with the largest concentration of \ce{CH4} (13\%) at the surface of the planet. The pressure variation between the modelled planets shown in the figure is not significant at lower altitudes but grows as the altitude increases. 

Unsurprisingly, planets with the large concentrations of \ce{CO2} and \ce{CH4} have hotter surface temperatures than on Modern Earth. However, the thermal inversion observed in the Modern Earth's stratosphere due to absorption of ultraviolet radiation by ozone (at around 50km altitude) is significantly affected by higher concentrations of \ce{CH4} and \ce{CO2}. In the case of higher \ce{CO2} concentration, although the surface is warmer due to the increased greenhouse effect, \ce{CO2} is also better able to cool in the infrared at altitudes above 30kms.

To gain a holistic view of the space of atmospheres simulated, the temperatures and pressures at the planetary surfaces were extracted, and the distribution of these atmospheres is shown in Fig.~\ref{fig:atmdist}. Since the distribution is sensitive to the scan parameters employed in the search, only limited conclusions are possible with the data collected here. Among the simulated atmospheres, there are three ``islands'' of atmospheres that can be identified in Fig.~\ref{fig:atmdist}. The bottom left-most of these contained the atmosphere corresponding to present-day Earth (average surface temperature of 15\textsuperscript{$\circ$}C and pressure of 1.02~atm). These islands are probably more a facet of the parameter space exploration strategy than indicative of planetary regimes.

\begin{figure*}[!bht]
    \centering
    \subfigure[]{ \includegraphics[width=0.31\hsize]{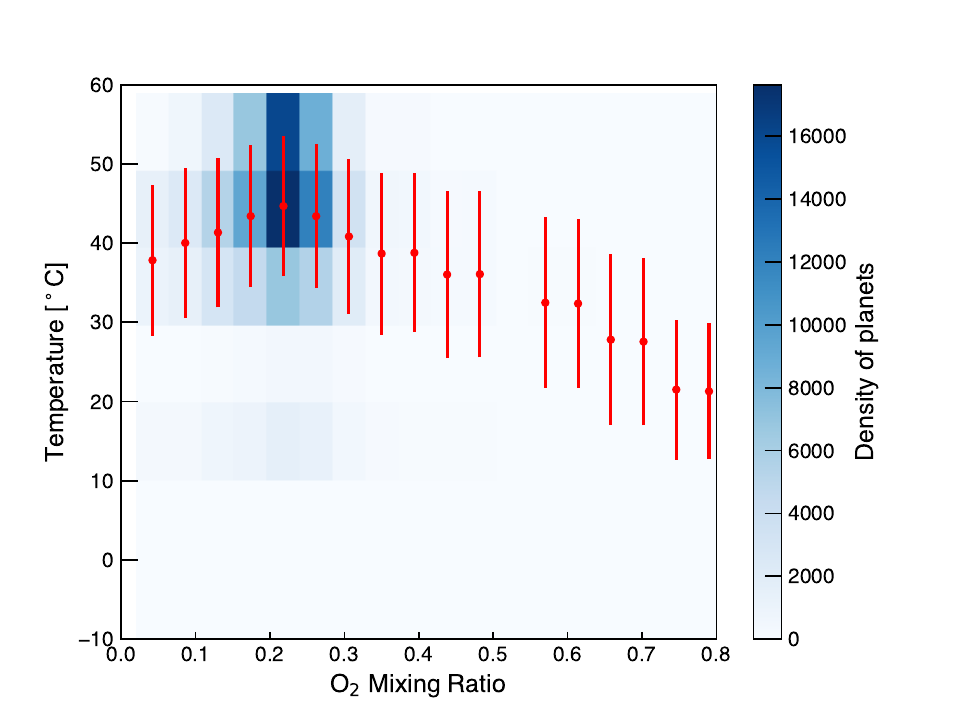} }
    \subfigure[]{ \includegraphics[width=0.31\hsize]{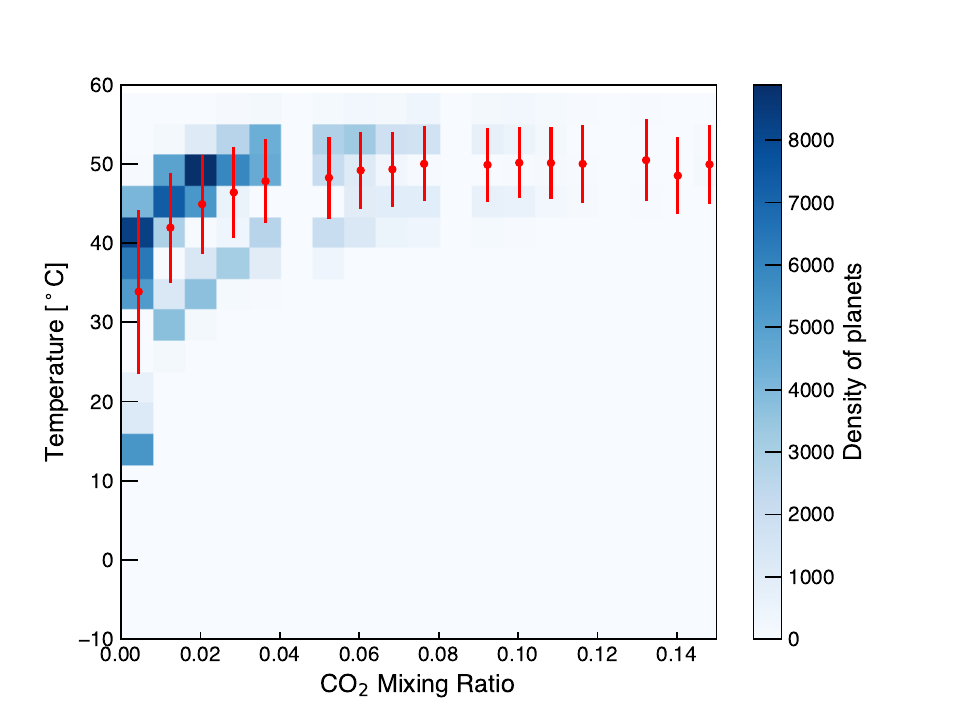} }
    \subfigure[]{ \includegraphics[width=0.31\hsize]{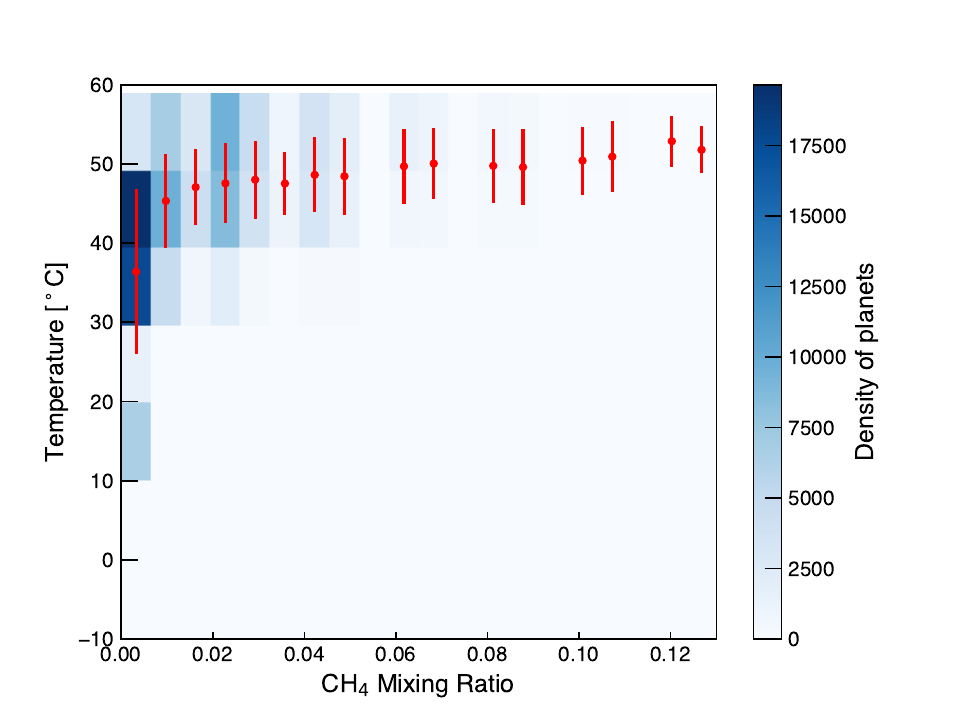} }
    \caption{2-D histograms (heatmaps) of the density of scanned atmospheres in the surface temperature versus gas mixing ratio plane, overlain with the profile histogram of atmosphere temperatures as a function of the gas mixing ratio (\ce{O2}-a, \ce{CO2}-b, \ce{CH4}-b). Each bin of gas mixing ratio contains many atmospheres, with all the combinations of other gasses that were simulated. Red points show the mean of the temperatures of the atmospheres in each bin, and the error bars show the standard deviation. As the heatmaps and the profile histograms depend significantly on the priors applied to the atmosphere scan and the number of scanned atmospheres, limited interpretation is possible with the current dataset and plots here only demonstrate the concept.}
    \label{fig:profiles}
\end{figure*}

Figure~\ref{fig:profiles} shows 2-D heatmaps and profile histograms for the \ce{O2} (\ref{fig:profiles}a) and \ce{CO2} (\ref{fig:profiles}b) concentrations versus temperature. The heatmaps are binned, and each bin shows the number of atmospheres generated as a function of temperature and the gas mixing ratio, with darker regions indicating a greater proportion of atmospheres in a given histogram bin. The profile histograms (red bars) show the average temperature for all the atmospheres in that particular range of gas mixing ratio; for example, the first red point on Fig.~\ref{fig:profiles}a corresponds to the average of all the atmospheres with \ce{O2} mixing ratio between 0.00--0.05 (regardless of the concentrations of other gases). The red point shows the mean of the temperatures of the atmospheres, and the error bars indicate the standard deviation. 

Plots such as Fig.~\ref{fig:profiles} offer a simple way of determining the surface temperature of a planet to first-order. Such an approximation would be particularly valuable where remote characterisation has only been able to constraint the abundances of some gases. For example, based on our current dataset, if we were to find that a planet had an \ce{O2} mixing ratio between 0.35--0.40, then there would be a 68\% chance that the surface temperature of that planet is in the range 30--50\,$^\circ$C -- a potentially useful result given that liquid water on the surface of a planet may be an indicator for life \citep{Chopra2016,Lineweaver2018}. Further constraints on the temperatures could be provided by a concordance of results from other gases. However, to infer realistic surface temperatures, we would need to simulate a representative set of all possible exoplanetary atmospheres and expand our current dataset.

\section{Future Directions}
\begin{enumerate}
\item This work set the stellar parameters to that of the Sun and an Earth-sized planet at 1AU from the host star. The work could be expanded to include M and K stars which are of particular interest to exoplanet habitability studies, and model a range of planetary sizes, insolation and obliquities. While we have used a relatively simple 1-D model in our study, the batch-processing framework developed to utilise cloud computing is sufficiently flexible to enable more recently developed and complex 3D-GCM models such as ROCKE-3D \citep{Way2017}, ExoCAM \citep{Wolf2018}, LMD-Generic \citep{Wordsworth2011, Turbet2018}, and the MET Office Unified Model \citep{Boutle2017} to be run at scale and conduct parameter sweeps. Such a grid of models, potentially validated with data from future exoplanet observations, could help estimate the statistical distributions of habitable zones.

\item Large collections of atmosphere models are valuable as synthetic training datasets for machine learning applications in the exoplanet science domains. For example, a neural network model trained on various stellar types, planetary radii, planet-star distances, and atmospheric compositions would reduce the need to run resource-intensive models. Similarly, ML-based atmospheric retrieval frameworks \citep{Soboczenski2018,Cobb2019} used to determine an exoplanetary atmosphere’s temperature structure and composition from an observed spectrum, can benefit from the large repository to atmospheric models to efficiently generate training spectra libraries.

\item In this study, we do not attempt to infer distribution of bio-masses and/or metabolisms capable of sustaining and co-existing with the surface gases fluxes of the modelled atmospheres. However, future simulations that couple planetary atmosphere models such as \ATMOS to biogeochemical models in a similar manner to \citet{Kharecha2005} and \citet{Gebauer2017}, could enable characterisation of the potential role of biology in regulating planetary atmospheres \citep{Harding2010,Lenton2018,Lyons2015}. 

Such efforts would lead to more nuanced context-dependent interpretations of habitability parameters such as surface temperatures, photon and redox free energy availability for different classes of planetary systems \citep{Lineweaver2018,Lenardic2021} and assessments of potential biosignatures in exoplanetary atmospheres.
\end{enumerate}

\section{Conclusions}
A set of \numatmospheres{} explanatory atmospheres have been simulated using a new framework, \pyatmos. For each simulated atmosphere, the temperature, pressure, gas concentration and gas flux were calculated as a function of altitude, ranging from 0--80\,km. The resulting dataset is much larger than was available before to the exoplanet habitability community. Using the computational framework, future work could facilitate statistical inferences on quantities such as an exoplanet's surface temperature and pressure based on more readily measurable quantities such as gas concentrations in a planetary atmosphere.

\bibliographystyle{cas-model2-names}

\bibliography{references}

\end{document}